\documentstyle[aps,epsf,preprint,graphicx]{revtex}
\begin{document}
\tightenlines
\title{Gauge invariant metric perturbations during reheating}
\author{Mauricio Bellini\footnote{E-mail address: mbellini@mdp.edu.ar ---
C\'atedra Patrimonial de CONACYT.}}
\address{Instituto de F\'{\i}sica y Matem\'aticas, 
Universidad Michoacana de San Nicol\'as de Hidalgo,\\
AP:2-82, (58041) Morelia, Michoac\'an, M\'exico.}
\maketitle
\begin{abstract}
The possible amplification of gauge invariant
metric fluctuations in the infrared
sector are very important during reheating stage of inflation.
In this stage the inflaton field oscillates around the minimum
of the scalar potential. The evolution for the super Hubble scales
gauge invariant metric fluctuations can be studied by means of
the Bardeen parameter. For a massive scalar field with
potential $V(\Phi_c) =
{m^2 \over 2} \Phi^2_c+\Lambda$ with nonzero cosmological constant.
I find that (in the reheating regime and for super Hubble scales)
the Sasaki-Mukhanov parameter oscillates with amplitude
constant such that there is not amplification of $Q$ during reheating.
\end{abstract}
\vskip 2cm
\noindent
PACS number(s): 98.80.Cq\\

\vskip 2cm
Postulating a period of nearly exponential growth in the primordial
universe, inflationary cosmology\cite{Guth} solved many
problems which plagued previous models of the big bang.
After inflation, the universe is dominated by the inflation,
the scalar field whose evolution controls the
dynamics of the inflationary era. The process for which
the inflaton energy density
is then converted into thermalized matter, is knowed as reheating\cite{reheat}.
The issue of gauge invariance becomes critical when we attempt
to analyze how the scalar metric perturbations produced in the very
early universe influence of a globally flat isotropic and
homogeneous universe\cite{Bardeen}.
The evolution of gauge invariant metric perturbations during inflation
have been well studied\cite{Gon}.
This allows to formulate the problem of the
amplitude for the scalar metric perturbations on the evolution
of the background Friedmann-Robertson-Walker (FRW)
universe in a coordinate - independent
manner at every moment in time.
Parametric resonance instability occurs during a reheating
period when the inflaton oscillates around the minimum of the
scalar potential\cite{Tras,Koda,Basset}.
Since the gravitational perturbation is coupled
to the inflaton field by the Einstein equation, it may also
experience parametric amplification during this stage.
The parametric resonance instabilities have important
consequences for cosmology. They will lead to a reheating
temperature which can be much larger than would
be obtained by calculating the
efficiency of reheating using perturbation theory and could
have important implications for
grand-unification-scale baryogenesis\cite{Riotto},
the production of supermassive dark matter\cite{Chung} or the formation
of topological defects\cite{Kasuya}.

In this work I study the metric inhomogeneities during reheating
for gauge invariant metric perturbations\cite{Bardeen} by means of the
Sasaki-Mukhanov variable\cite{SM} on super Hubble scales
for a massive scalar field with potential
$V(\Phi_c)={m^2 \over 2} \Phi^2_c + \Lambda$ and nonzero cosmological
constant and $\Phi_c=+ {3 M_p H_0\over \pi m}$.
Since the results do not depend on the gauge, the perturbed
globally flat
isotropic and homogeneous universe is well described by\cite{Bardeen}
\begin{equation}\label{m}
ds^2 = (1+2\psi) \  dt^2 - a^2(t) (1-2\Phi) \  dx^2,
\end{equation}
where $a$ is the scale factor of the universe and $\psi$ and $\Phi$
the perturbations of the metric. I will consider the particular
case where the tensor
$T_{ij}$ is diagonal, i.e., for $\Phi = \psi$\cite{4a}.
Linearizing the Einstein
equations in terms of the matter and metric fluctuations
$\phi$ and $\Phi$, one obtains the system of
differential equations for $\phi$ and $\Phi$\cite{Linde}
\begin{eqnarray}
 \ddot\Phi &+& \left(\frac{\dot a}{a}
- 2 \frac{\ddot\phi_c}{\dot\phi_c} \right)
\dot \Phi - \frac{1}{a^2} \nabla^2 \Phi   
+2\left[
\frac{\ddot a}{a} - \left(\frac{\dot a}{a}\right)^2 - \frac{\dot a}{a}
\frac{\ddot\phi_c}{\dot\phi_c}\right] \Phi =0, \label{1}\\
\frac{1}{a}& \frac{d}{dt}& \left( a \Phi \right)_{,\beta} =
\frac{4\pi}{M^2_p} \left(\dot\phi_c \phi\right)_{,\beta} , \\
\ddot\phi& +& 3 \frac{\dot a}{a} \dot\phi -
\frac{1}{a^2} \nabla^2 \phi + V''(\phi_c) \phi 
+ 2 V'(\phi_c) \Phi- 4 \dot\phi_c \dot\Phi =0.
\end{eqnarray}
Here, the dynamics of the spatially homogeneous field
$\phi_c$ being described by the equations\cite{BCMS}
\begin{eqnarray}
&& \ddot\phi_c + 3 \frac{\dot a}{a} \dot\phi_c + V'(\phi_c) = 0, \\
&& \dot \phi_c = - \frac{M^2_p}{4 \pi} H'_c(\phi_c),
\end{eqnarray}
where the prime denotes the derivative with respect to $\phi_c$ and
$H_c(\phi_c) \equiv {\dot a\over a}$.

In the case of adiabatic perturbations with scales outside the Hubble
radius, it is convenient to work with the Bardeen parameter
\begin{equation}
\zeta = \frac{2}{3} \frac{\Phi + H^{-1}_c \dot\Phi}{1+\omega} + \Phi,
\end{equation}
where the overdot denotes the time derivative and $\omega = p/\rho$
is the ratio of the pressure to the density of the background.
In the $k \rightarrow 0$ limit, the Bardeen parameter
satisfies\cite{Bardeen1,4a}
\begin{equation}
\frac{3}{2}(1+\omega) H_c \dot\zeta =0,
\end{equation}
which, in the reheating stage holds the following
condition\cite{4a}
\begin{equation}
\dot\phi^2_c \dot\zeta =0.
\end{equation}
During the slow roll regime the eq. (\ref{1}) is well behaved, but
during the reheating period the field $\phi_c$ oscillates and
the coefficients in eq. (\ref{1}) has singularities, which
can be removed
by means of the Sasaki-Mukhanov variable $Q = \phi+{\dot\phi_c \over
H_c} \Phi$. Hence, from eq. (\ref{1}), the modes for $Q$ satisfies
\cite{Mu}
\begin{equation}\label{Q}
\ddot Q_k + 3 H_c \dot Q_k + \left[ V'' +\frac{k^2}{a^2} +
2 \frac{d}{dt}\left(\frac{\dot H_c}{H_c} + 3 H_c \right)\right] Q_k =0.
\end{equation}
The parameter $Q$ is related with the Bardeen one by the
following relation\cite{Finelli}
\begin{equation}\label{xi}
\zeta = \frac{H_c}{\dot\phi_c} Q.
\end{equation}
The structure of the eq. (\ref{Q}) can be simplified with the
map $Q = e^{-3/2 \int H_c dt} \  P$, such that the equation for
the modes of $P$ is
\begin{equation}\label{P}
\ddot P_k + \left[ \frac{k^2}{a^2}
+ V'' + \frac{9}{2}\dot H_c
-\frac{9}{4} H^2_c 
+2 \frac{\ddot H_c}{H_c} - 2 \left(\frac{\dot H_c}{H_c}
\right)^2 \right] P_k =0,
\end{equation}
which can be written as $
\ddot P_k + \left[{k^2\over a^2} - {k^2_0 \over a^2}\right] P_k = 0$,
where $\mu^2(t)={k^2_0 \over a^2} $ is an effective parameter
of mass given by
\begin{equation}
\mu^2(t) =
2 \left(\frac{\dot H_c}{H_c}
\right)^2 - V'' - \frac{9}{2}\dot H_c
-\frac{9}{4} H^2_c - 2 \frac{\ddot H_c}{H_c} .
\end{equation}
In the infrared sector one takes ${k^2\over a^2} \ll {k^2_0 \over a^2}$,
where $k_0$ is the time dependent wavenumber which separates the
infrared and ultraviolet sectors.
During inflation, for
a given scalar potential $V(\phi_c)$, the dynamics of the
Hubble parameter $H_c(\phi_c)$ being described by
the Friedmann equation (see, for example,\cite{prd96})
\begin{equation}\label{V1}
V(\phi_c) = \frac{3 M^2_p}{8\pi} \left[ H^2_c - \frac{M^2_p}{12 \pi}
\left(\frac{\dot H_c}{\dot\phi_c}\right)^2 \right].
\end{equation}
In the reheating period the second term inside the brackets cannot
be neglected because the slow roll conditions are violated.

Now I consider the scalar potential $V(\Phi_c) = {m^2\over 2}
\Phi^2_c + \Lambda$,
where $\Lambda$ is some negative cosmological constant and
$\Phi_c=\phi_c+{3 M_p H_0 \over \pi m}$.
If $\Phi_c(t) = {2\over 5} M_p \ {\rm cos}(\omega \tau)$ [here $\tau =
t-t(\phi_c=0)$], one obtains
$H_c=4 \sqrt{{\pi\over 75}} m \  {\rm cos}(\omega \tau)+H_0$, $V''=m^2$
and $\dot H_c = - 4
\sqrt{{\pi\over 75}} m \  \omega \  {\rm sin}(\omega \tau)$.
Here, $\omega$ is
the oscillation frequency of the inflaton field and $H_0 \equiv
H_c(\phi_c=0)$.
From eq.
(\ref{V1}) one obtains the cosmological constant
\begin{equation}
\Lambda = -\frac{M^2_p}{2\pi} \left[ H^2_0\left(\frac{36-3\pi}{4\pi}\right)
+ \frac{m^2}{12\pi}\right].
\end{equation}
This case was studied by Finelli and Branderberger\cite{Finelli} 
but without cosmological
constant. In this paper I am interested in the study of possible
consequences in the evolution of super Hubble 
metric perturbations during reheating when one
take into account a nonzero cosmological constant.
During reheating $0 < H_c < m$, so that
$m > H_0 > {4\over 5} m$  and $\Lambda $ can take
negative values near the minimum of the potential.
The effective parameter of mass is
\begin{eqnarray}
\mu^2(\tau) &=& 2
\left(\frac{4 \sqrt{{\pi\over 75}} m \omega {\rm sin}(\omega\tau)
}{4 \sqrt{{\pi\over 75}} m {\rm cos}(\omega\tau) +H_0}\right)^2 +
18 \sqrt{{\pi\over 75}}
\left[{\rm sin}(\omega\tau)\right.\nonumber \\
&-& \left.{\rm cos}(\omega\tau)\right]
- \frac{12\pi}{25} m^2 {\rm cos}^2(\omega\tau) 
- \frac{9}{4}
H^2_0 + \frac{8 \sqrt{{\pi\over 75}} m \omega^2 {\rm cos}(\omega\tau)}{
4\sqrt{{\pi\over 75}} m {\rm cos}(\omega\tau) +H_0}.\label{mu}
\end{eqnarray}
Since in the infrared sector ${k^2 \over a^2} \ll {k^2_0 \over
a^2}$, the solution for $P_k$
is given with a very good approximation for super Hubble scales by
the equation $\ddot P_k - \mu^2(\tau) P_k =0$.
For inflation take place one requires that $\mu^2(\tau)\ge 0$.
The conditions required are (for $\omega <1$; $m =10^{-5} M_p$)
\begin{eqnarray}
&& \frac{\omega}{m} >1, \\
&& H_0 - 4 \sqrt{\frac{\pi}{75}} \  m  \ll 10^{-1} \  M_p.
\end{eqnarray}
Figure 1 shows the numerical solution for
the evolution of $P \  e^{-{3\over 2} H_0 t} \equiv
Q \  e^{{3 m \over 2 \omega} {\rm sin}(\omega\tau)}$ for
$\tau \ge 0$ and $\omega = 10^{-3} \  M_p$. Note that the amplitude
of $Q \  e^{{3 m \over 2 \omega} {\rm sin}(\omega\tau)}$ increases
with time. This means that $Q$ also increases with time.
Note that in this case $\mu^2<0$ and
inflation do not take place. 

Figure 2 shows
$P \  e^{-{3 \over 2} H_0 t}$ for $\omega = 10^{-5} \  M_p=m$. Note
that $Q \  e^{{3 m \over 2 \omega} {\rm sin}(\omega\tau)}$ oscillates
with amplitude constant and effective frequency $\omega_{eff} \simeq
6 \times 10^{-3} \  M_p$. Hence, the solution for the Sasaki-Mukhanov
variable will be a coupled oscillator with frequencies $\omega =
10^{-5} \  M_p$ and $\omega_{eff}\simeq 6 \times 10^{-3} \  M_p$.
In this
case $\mu^2 >0$ and $\mu^2 \gg {k^2 \over a^2}$ in the IR sector.

Finally, from eq. (\ref{xi}) one obtains the Bardeen parameter
for super Hubble scales during the reheating era
\begin{equation}\label{iu}
\zeta = -\frac{5}{2} \left[\frac{4 \sqrt{{\pi\over 75}} m {\rm cos}(\omega\tau)
+H_0}{M_p \omega {\rm sin}(\omega\tau)}\right] \  Q.
\end{equation}
Note that the denominator in eq. (\ref{iu}) has singular points
in $\omega\tau=n\pi$ ($n=0,1,2,...$). Hence, the Bardeen parameter
indeterminates for $\omega\tau = n\pi$. These singularities in
$\zeta$ correspond to the the points where $\dot\phi_c=0$.
This result disagrees with
other results of a previous work\cite{Lin}, where the authors founded
that $\zeta$ oscillates with constant amplitude during reheating.

The results here obtained for $Q$ are very similar with the Finelli
and Branderberger\cite{Finelli} (i.e., for $\mu^2 >0$) because there
is not growth of the Sasaki-Mukhanov variable for super Hubble scales
during reheating. The only difference is that here $Q$ oscillates around
a constant value. This oscillation should be a consequence that ---
instead Finelli and Branderberger have made to simplify the equation of
motion for $P_k$ (\ref{P}) ---
I worked without perform a time average in the effective parameter
of mass [see eq. (\ref{mu})]. The calculation without
a time average leads to an oscillating $\mu^2$ which is the responsible
for the oscillation of $Q$.
A similar result, but when super
Hubble non-adiabatic pressure perturbation is negligible was obtained
recently by Wands {\em et. al.}\cite{Wands}.

\begin{center}
\begin{figure}
\noindent
\includegraphics[width=10cm, height=12cm]{c:/tex/fgi4.bmp}
\vspace{-0cm} \noindent \caption{\label{fig1} Evolution for
$P/a^{3/2}$ as a function of $\tau$, for $H_0 = 0.4 m$, $m=10^{-5}
\  M_p$ and $\omega=10^{-3} \  M_p$. The amplitude of $P/a^{3/2}$
increases with time in the ultraviolet sector. Note that we have
used Planckian unities.}
\end{figure}
\end{center}
\newpage
\begin{center}
\begin{figure}
\noindent
\includegraphics[width=10cm, height=12cm]{c:/tex/fgi41.bmp}
\vspace{-0cm} \noindent \caption{\label{fig2} Evolution for
$P/a^{3/2}$ as a function of $\tau$, for $H_0 = 0.4 m$, $m=10^{-5}
\  M_p$ and $\omega=10^{-5} \  M_p$. The amplitude of $P/a^{3/2}$
remains constant. Note that $P/a^{3/2}$ oscillates with effective
frequency $\omega_{eff} \simeq 6 \times 10^{-3} \  M_p$.}
\end{figure}
\end{center}

\end{document}